\documentclass[usenatbib]{basi}
\usepackage[T1]{fontenc}
\usepackage[british]{babel}
\usepackage[varg]{txfonts}
\usepackage{rotating}
\usepackage{dcolumn}

\def\deg{$^\circ$}
\def\arcsec{"}
\def\aap{A\&A}

\def\apj{ApJ}
\def\apjs{ApJ Sup. Ser.}
\def\aj{AJ}
\def\apjl{ApJL}

\def\nat{Nature}

\def\mnras{MNRAS}
\def\pasp{PASP}
\def\araa{An.Rev. A\&A}

\def\kms{${\rm km}\cdot{\rm s}^{-1}$\,}

\def\rsun{R$_{\odot}$\,}
\def\msun{M$_{\odot}$\,}

\begin{document}
\title[BASI sample paper]{Interferometric studies of novae in
the infrared}
\author[O.~Chesneau \& D.P.K.~Banerjee]%
       {O.~Chesneau$^1$\thanks{email: \texttt{Olivier.Chesneau@oca.eu}},
       and D.P.K.~Banerjee$^{2}$ \thanks{email: \texttt{orion@prl.res.in}}\\
       $^1$Laboratoire Lagrange, UMR7293, Université de Nice Sophia-Antipolis, CNRS, Observatoire de la Côte d'Azur, 06300 Nice, France\\
       $^2$Physical Research Laboratory, Navrangpura, Ahmedabad, Gujarat, India}

\pubyear{2012}
\volume{00}
\pagerange{\pageref{firstpage}--\pageref{lastpage}}
%\status{submitted}

\date{Received --- ; accepted ---}

\maketitle
%------------------------------------------------------------------------------%
% abstract and keywords                                                        %
%------------------------------------------------------------------------------%
\label{firstpage}

\begin{abstract}
 The Interferometric studies of novae in the optical and near-infrared is a nascent but fast emerging field which has begun to provide new and invaluable insights into the nova phenomenon. This is particularly so in the  early stages of the eruption when all the relevant  physical phenomena are on the scale of milli-arcseconds and thus are amenable to be studied only by interferometric techniques. In this review the instruments and arrays involved in this domain of work   are briefly described, followed by a description of the major results obtained so far. A discussion is made of the  physical aspects,  where the application of interferometric techniques, can bring the most valuable information. Finally,  prospects for the near future are discussed.
\end{abstract}


%------------------------------------------------------------------------------%
% main text of the paper, using \section, \subsection, \subsubsection          %
%------------------------------------------------------------------------------%
\section{Introduction}\label{s:intro}

 A classical nova eruption results from a
thermonuclear runaway (TNR) on the surface of a white dwarf (WD) that is accreting material from a companion star in a close binary system. The accreted hydrogen rich material gradually forms a layer on the WD's surface with the mass of the layer growing with time. As the accreted matter is  compressed and heated by the gravity of the WD,  the critical temperature and pressure  needed for the TNR  to commence are  reached in the degenerate material at the base of the accreted layer thereby starting  the nova eruption.  Observationally, the outburst is accompanied by a large brightening, generally with an amplitude of 7 to 15 magnitudes above the brightness of the object in quiescence (Warner 2008). Subsequently, the light curve decays and the rate of decline has been used to estimate the absolute magnitude and distance to the object through empirical relations relating the maximum brightness to the rate of decline (MMRD) relations (e.g. Della Valle \& Livio 1995;  also see Warner 2008 for a summary of other relationships). The key parameters of these relations are the $t_{2}$  and $t_{3}$ times i.e. the times for the brightness to decline by 2 and  3 magnitudes respectively from peak brightness, usually in the visible band. Such relations are derived from samples of novae in the same external galaxy which thus eliminates uncertainties in distance to individual novae and thereby provides a reasonably  effective  calibration scheme for  estimating distances. The nature and evolution of the light curves can be diverse -  the recent compilation and  classification scheme  of novae light curves by \citet{2010AJ....140...34S} illustrates this.

At outburst, the spectral energy distribution of a nova is generally found from several studies to be adequately characterized by a blackbody with an effective temperature between 6000 - 10000K (i.e. resembling the SED of a star of F or A spectral type; Gehrz 1988). This blackbody behavior is attributed to the nova's pseudophotosphere or equivalently the  surface  of the optically thick expanding fireball within which is contained the dense, initially ejected material.  A nova spectroscopically shows at this stage, both in the optical and IR, several emission lines with prominent P Cygni features. The absorption component of the P Cygni feature can dominate over the emission component in the very early stages. With time, the lines gradually change to being purely in emission and the degree of excitation and ionization is  seen to gradually increase with time culminating in many cases in a  coronal phase characterized by emission lines arising from  highly ionized species.  The development and evolution of the optical spectra are described in Williams (1992) wherein it is shown  that the spectra can basically be divided into two broad classes viz the FeII and the He/N novae depending on the species  which shows the next-most prominent lines (FeII or He/N lines) after hydrogen. Many novae, especially those of the FeII type, can produce significant amounts of dust later during their evolution. At such stages the optical light curve generally manifests a sharp decline due to obscuration of the central source by the dust; this is accompanied by a sharp increase in the IR emission because of the dust. The dip in the light curve  may not always happen - novae are known to produce optically thin dust shells too.  All aspects of the outburst, from the initial fireball phase, followed by various stages in the emission line phase,  and  a dust formation phase if it occurs can be studied with interferometry, both in the near and mid-IR. It is important to realize that the detailed geometry of the first phases of a nova in outburst remains virtually unexplored.

The central binary in a nova system has a WD primary, of either the CO or ONe type. The WD accretes material from a low mass main sequence companion, orbiting with a period of P$\sim$1.4-8h, or in the case of fewer systems from a giant filling its Roche Lobe with longer periods. The extreme case being the (rare) symbiotic systems in which the period reaches several years. Classical Novae have typical inter-outburst timescales of the order 10$^4$ years as predicted from theoretical models \citep*[for. e.g a large grid of nova models is given in][]{2005ApJ...623..398Y}. However unlike classical novae, the inter-outburst timescale for recurrent novae can be as low as a decade. The short recurrence periods of RNe require high mass WD accretors and relatively high accretion rates.  It has been argued that recurrent novae systems are the progenitors of Type Ia supernovae, since the WD likely gains mass
with every outburst coupled to the fact that the short recurrence time indicates a  WD  very close to the Chandrasekhar limit. A few words on recurrent novae may be in order since two of our major interferometric results concern them and it is necessary to understand their environments when interpreting data. The Recurrent Novae can be divided into several groups.

{\it The RS Oph/T CrB group} which have  red giant secondaries, consequently long orbital
periods (several hundreds of  days) and which show strong
evidence for the interaction of the ejecta with the pre-existing circumstellar wind
of the red giant.
The  {\it U Sco group} comprise of  systems containing
an evolved main sequence or sub-giant secondary with an orbital period much
more similar to that in CNe (of the of order of hours to a day; U Sco for example has $P$$_{orb}$ = 29.53h), a rapid optical decline
and extremely high ejection velocities ($v_{ej}$ $\sim$ 5000-10 000 km.$s^{-1}$). The $t_{2}$  and $t_{3}$ times are of the  order of a few days, which is generally too short a time  required to prepare, activate and undertake optical interferometry observations. The rapid decline rates inhibits interferometric observations considerably compared to more conventional techniques such as photometry or spectroscopy.
The {\it T Pyx group} again comprises of short orbital period systems. They show a very heterogeneous set of moderately fast to slow optical light curve declines.

Nova expansion parallax measurements are used in studies of individual novae and in statistical studies such as MMRD relations. Such distance estimates compare the observed angular size of a nova shell with its linear size which is inferred from expansion velocities obtained from line profiles coupled with the known time over which the expansion has taken place. The angular size is most of the time inferred from photometric measurements. Since the expansion rate of nova remnants is  small as projected on the sky,  the expansion is  mostly studied over several years  or tens of years. The facility that provided significant improvement in this aspect was the HST with its high resolution imaging capability, although long slit spectroscopy from the ground has helped considerably in inferring the kinematics, hence the true 3D shape of the structures \citep{2011PhDT.........1R, 2009AJ....138.1541M, 2007ApJ...665L..63B}. Usual features seen in classical novae include polar blobs and polar rings/shells with tropical and equatorial bands. Investigating the nebular remnants of novae and studying the shaping processes acting on the ejecta has wide implications for example in our understanding of the shaping of proto-Planetary Nebulae.

We have presented a brief introduction of the nova phenomenon here, largely intended for those working in the field of interferometry. The information purveyed here is therefore mostly related to aspects  related  to this field. However the reader may find the present article more self-contained,  if  a few other sources covering other aspects of the nova phenomenon are mentioned. These are contained in several articles collated in the  book on Classical Novae  by \citet{2008clno.book.....B} and in the book on Cataclysmic Variables by Warner (2003). Observationally, of particular interest in the former book may be the articles  related to the historical perspective by Duerbeck (2008), an overview of the observational properties of novae by Warner (2008) and that related to the shape and shaping of resolved novae remnants by O'Brien and Bode (2008). Articles with a broad overview, to mention a few, related to the infrared development, dust formation and the strikingly distinctive elemental abundances seen in novae can be found in  \citet{1988ARA&A..26..377G}, \citet{1997ApJ...479L..55J} and \citet{1998PASP..110....3G}.

Let us assume as an illustrative example, that  a cataclysmic variable system is being observed at a distance of one kilo-parsec. At this distance, one Astronomical Unit (1 A.U) will subtend an angle of one milli-arcsecond (mas) in the sky. Let us further assume, that one mas is the typical resolution of most optical interferometers\footnote{The term optical interferometry is used in a general sense, to encompass studies in both the visible as well as in the infrared regime}.  The  WD and the secondary which have periods ranging from 1h to 8h, are then separated by less than 0.01AU (i.e. 0.01mas) at this distance and cannot be resolved. They are additionally intrinsically faint, and cannot be studied during quiescence by optical interferometry. To summarize, the evolution of a nova outburst can be divided into several stages (see Fig. 2). \newline
- after a nova explosion sets in, the optically thick photosphere expands greatly up to $\leq$100\rsun, and the companion star is engulfed deep inside the photosphere. The angle subtended by the source is between 0.01 mas up to 1\,mas.  \newline
- after the maximum expansion, the photospheric radius shrinks with time and free-free emission dominates the flux at relatively longer wavelengths. The near-IR nova is best described by a two component model, with a compact, nearly unresolved central source and the expanding ejecta. The
mean velocity of the ejecta appears related to the speed class of the CV (McLaughlin
1960). Slow to very slow novae exhibit ejecta at velocities ranging from 500 to 1000\kms\, while the ejecta  from fast novae travel more rapidly, with velocities ranging from 2000 to 3000\kms and sometimes even faster. \newline
- the companion eventually emerges from the WD photosphere and an accretion disk may appear or be reestablished again; the photosphere further shrinks to a size of $\leq$0.1\rsun\ and  the optically thick wind eventually stops; the ejecta get fully decoupled from the core. At this stage the flux from the source has decrease much beyond the sensitivity limit of any interferometer.

Speed classes reflect the rate of decline from maximum light. An important point to consider  is that material moving at a greater velocity will fade quicker. Fast novae therefore, even the brightest ones, cannot be observed for long by optical interferometry given the current limits of  sensitivity of such instruments (to be discussed below).  There is therefore a large bias to observe and study the bright and slowly evolving  novae.

\begin{figure}
\centerline{\includegraphics[angle=270, width=10cm]{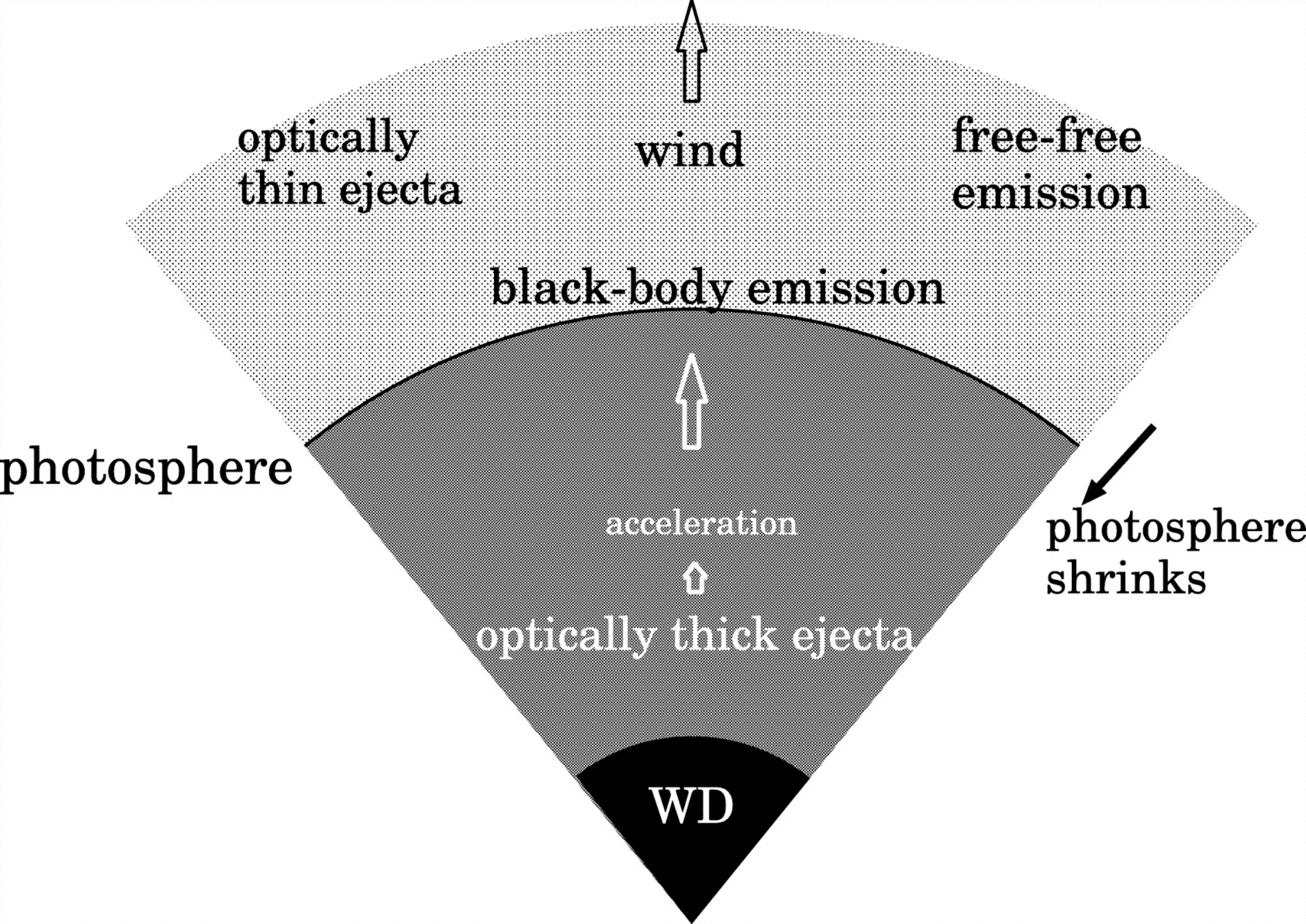}} %
\caption{Schematic configuration of a nova ejection model adopted from \citet{2006ApJS..167...59H}: A large part of the initial envelope mass is ejected by the winds, which are accelerated deep inside the photosphere. After the optical maximum, that is, after the maximum expansion of the photosphere, the photosphere begins to shrink, whereas the ejecta keep expanding. The optically thin layer emits free-free radiation at relatively longer wavelengths, while blackbody radiation from the photosphere dominates at shorter wavelengths.\label{f:kato}}
\end{figure}

\begin{figure}
\centerline{\includegraphics[width=10cm]{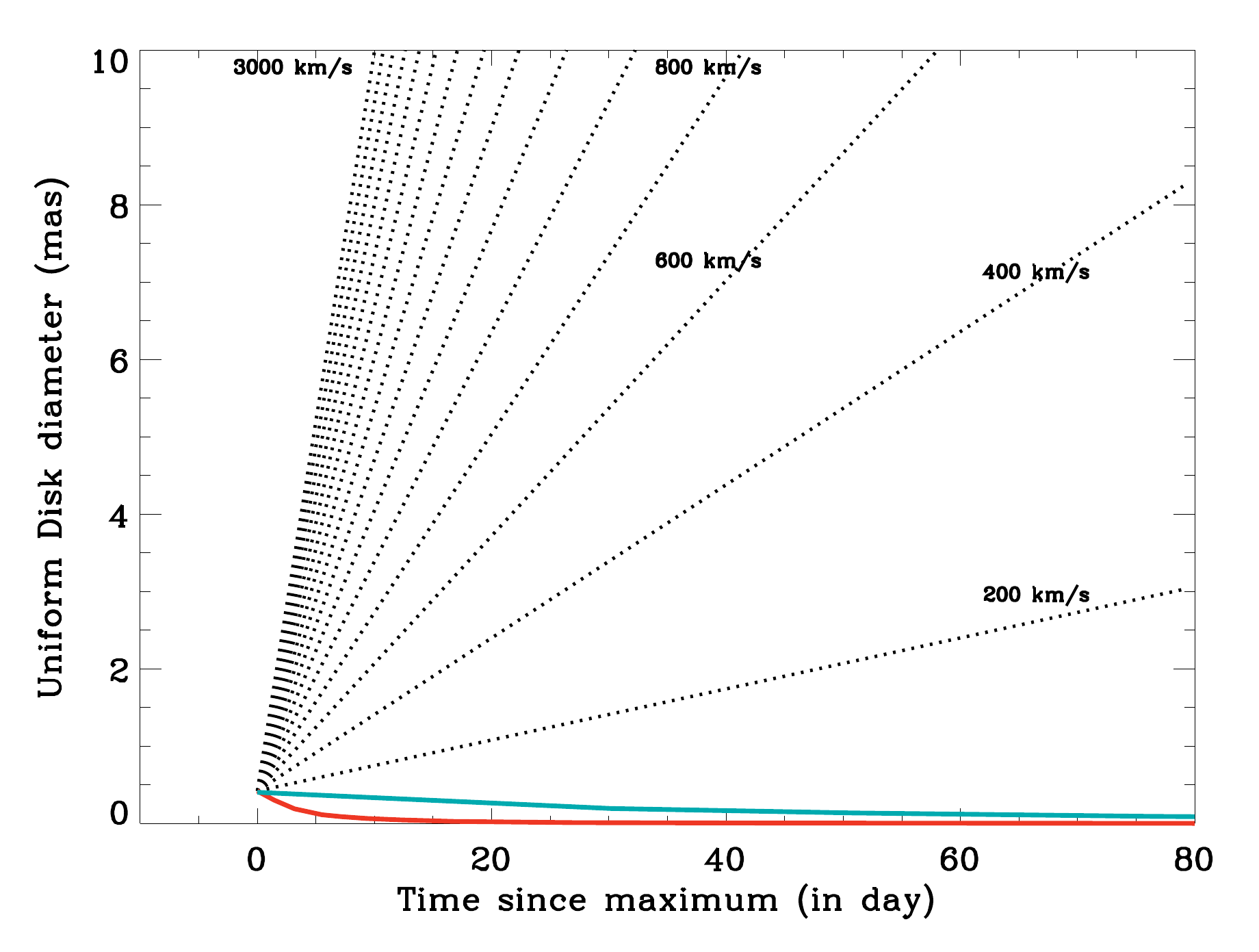}} %
\caption{Expansion curves for the ejecta of a nova located at 1 kpc. For comparison, the evolving size of the shrinking optically thick photosphere surrounding a 1.2\msun\ (red) and 0.6\msun (yellow) white dwarf are indicated. These latter curves are extracted from \citet{1997ApJS..113..121K}. \label{f:exp}}
\end{figure}

\section{Instruments}\label{s:instruments}

\subsection{The VLTI array}
The ESO Very Large Telescope Interferometer (VLTI) is located on Cerro Paranal,
Chile. This facility consists of the four fixed 8.2m Unit Telescopes
(UTs) and four 1.8m Auxiliary Telescopes (ATs) which can be moved to different locations
over an array of 30 stations. The VLTI offers two instruments viz. MIDI operating in the mid-IR and AMBER in the near-IR which are able to recombine the light of 2 or 3 telescopes respectively. Both instruments are well suited to study the first moments of a nova explosion. AMBER is able to resolve a spatially  bright near-IR source ($K$ $\leq$5-6 magnitudes, 2.2$\mu$m) of a few mas with a spectral resolving power of 1500. The source must be larger than $\sim$2mas and is over-resolved when it is larger than $\sim$30mas. It was successfully used  to observe the 2006 outburst of RS\,Oph, the 2011 outburst of T\,Pyx and the fast nova V5583\,Sgr which erupted in 2009. MIDI is more adapted to observe bright  mid-IR sources ($N$ $\leq$4 magnitudes, 8-13.5$\mu$m), especially those forming dust such as the slow nova V1280\,Sco, with a spatial resolution of the order of 10 mas. The source is over-resolved when larger than $\sim$100mas. Recently PIONIER, a new 4-telescope visitor instrument for the VLTI, that recombines light at low resolution in the $H$ band has also become operational and provided unprecedented snapshot observations of the recurrent nova T\,Pyx in outburst \citep{2011A&A...534L..11C}.

The VLTI is operated like any large facility. Telescope time for ESO telescopes at the La Silla/Paranal Observatory, including the VLTI,  is allocated twice a year in periods of 6 months. This is an important fact to mention since this is the only optical interferometer that is open to a large and general community, the others being operated by dedicated teams primarily with institutional interests. This also implies that for observing transient phenomena, Director's Discretionary Time Proposals (DDTs) need to be proposed using the same procedures as  for normal proposals. DDT proposals are reviewed by the Director's Discretionary Time Committee (DDTC). Note also that the observations of transients are best performed in service mode, implying that the observations must be performed with the modes officially offered to the community. There is also the Target of Opportunity (ToO) proposals that guarantee, and reserve in advance, an amount of observing time for transient phenomena. However, our experience is that the number of sources observable per semester is well below one, and the carry-over of a VLTI/ToO is too demanding in terms of efforts to be really useful and necessary. Furthermore, when a great deal of time has been allocated in the ToO mode,  the temptation is high  to use it, rather than let the time lapse,   on sources barely at the limiting magnitude of the instruments . This does not promote productive results and the DDT approach remains the best.

\subsection{The CHARA array}
The Center for High Angular Resolution Astronomy (CHARA) is an optical/interferometric array of six 1m telescopes located on Mount Wilson California, separated by distances up to 330 meters \citep{2005ApJ...628..453T}. CHARA offers many instruments capable of recombining visible and near-IR light from 2 to 6 telescopes with the great advantage of being able to conduct simultaneous observations with two sub-arrays composed of usually two and four telescopes. To date, CHARA observations of novae have been performed with the two-beam recombiner CLASSIC \citep[see Sect. on T\,Pyx]{2011A&A...534L..11C}, but the four-beam recombiner VEGA \citep{2009A&A...508.1073M} or the 4/6-beam recombiner MIRC \citep{2010SPIE.7734E..13M} have  great potential in this field (see Section\,8 on future prospects). Sources larger than 0.25mas can be resolved in the visible and are over-resolved when reaching about 8mas, translating to 1mas and $\sim$30mas in the K band.

When a potentially observable event occurs, a message is sent to the community of CHARA users, and observations are performed in a flexible manner, depending on the good will of the observers at the mountain and their collaborators. The observations are conducted from March to end of December, the winter period at Mt Wilson being relatively unproductive. The brightness limits under which the CHARA instruments observe in optimal conditions is 6-7 magnitudes in $K$ and $V$ at low spectral resolution, and  5-6 magnitudes  in $V$ with the VEGA instrument with a moderate spectral resolution of R=6000.

\subsection{Other optical interferometry facilities}

%The Navy Prototype Optical Interferometer (NPOI) recently renamed the NOI, is a mature facility that currently can combine arrays of 3 and 4 telescopes, with baselines reaching 90m.

The Palomar Testbed Interferometer (PTI) was a near-infrared, long-baseline
stellar interferometer located at Palomar Observatory in California that successfully observed the outbursts of various transient events including V838\,Monocerotis, V1663\,Aquilae and RS\,Ophiuchi until its closing in December 2008 \citep{2007ApJ...669.1150L, 2007ApJ...658..520L, 2005ApJ...622L.137L}. PTI had three 40 cm apertures that could be combined pair-wise to provide baselines up to 110 m, and recorded scientific data in the $K$ band.

The Keck I and Keck II telescopes can also work together as the Keck Interferometer with a 85 meter baseline. First fringes with the interferometer were obtained in March 2001 using the two Keck telescopes with their adaptive optics systems. The so-called Keck Interferometer Nuller (KIN) was involved in observations of RS Oph \citep{2008ApJ...677.1253B}. It was designed to null the mid-infrared emission from nearby stars so as to ease the measurement of faint circumstellar emission.

The Infrared Optical Telescope Array (IOTA) was a very productive Michelson stellar interferometer located on Mount Hopkins in southern Arizona. It operated with three 45 cm apertures that could be located at different stations on each arm of an L-shaped array (15 m X 35 m, maximum baseline 38 m). IOTA operated with 2 telescopes from 1995, and 3 telescopes from February 2002 to the end of June 2006. This instrument was also involved in the observations of RS Oph \citep{2006ApJ...647L.127M}.

\subsection{Supporting facilities: Mount Abu}
The 1.2m Infrared Telescope at Mount Abu, Rajasthan is  a facility in India \citep{2010ASInC...1..211A} specifically devoted to ground based infrared observations  due to the reasonably low amount of precipitable water vapour at the site \citep[1 to 2 mm during winter]{1995BASI...23...13D}. There is no optical interferometer at Mt Abu, but the telescope is mentioned because it provides in advance some crucial information regarding the photometry of the source, as also spectroscopic information, which greatly facilitates the planning and execution of the interferometric observations.

Optical interferometers are limited in terms of sensitivity or source brightness. Hence, before taking the decision to observe  a source  or deciding to submit a DDT proposal, it has to be ensured  that the source is bright enough and evolving favorably in  brightness before it can be observed. The experience at the VLTI is that a source can be observed comfortably if its  $V$ magnitude is brighter than 8 and the $K$ band magnitude brighter than 5. The results presented below were all secured when the sources were brighter than these thresholds. Hence, it is very important to observe the source in the $K$ band as soon as possible, and continue monitoring it  over the early evolution. The Mt Abu observatory is an  active near-infrared facility in the network of nova studies. Early exchanges of information on the near-IR status and evolution of an object coupled with  observations reported on RS\,Oph \citep{2006ApJ...653L.141D, 2006CBET..730....1D}, V1280\,Sco \citep{2008MNRAS.391.1874D, 2007CBET..864....1D} or T Pyx \citep{2011ATel.3297....1B} were used to plan the interferometric observations and then interpret them. One can also cite the case of V830\,Mon mentioned in this review, an object pursued in the past from Mt. Abu \citep{2002AA...395..161B, 2005ApJ...627L.141B} and  from where recent data has also been useful in planning for the interferometry.

\subsection{Instrumental prospects}
The VLTI is evolving and the first generation of instruments will soon be followed by a newer and much more powerful suite of instruments. These are PRIMA, GRAVITY and MATISSE which will offer considerably enhanced sensitivity and imaging capabilities.

A technological jump will occur with the advent of dual-fed instruments, capable of simultaneously observing the science target and a close-by astrometric reference. For instance, VLTI/PRIMA will feed the MIDI and AMBER instruments with a two beam input. By using internal metrology and a proper knowledge of the telescope array localization it will be possible to not only provide a measurement of the astrometric distance between the two objects, but also to record 'blindly' fringes of the science target by tracking the optical delay using the reference target. Similar instrumentation is in development for the Keck Interferometer also, the so-called KI/ASTRA.
The second generation VLTI/GRAVITY instrument is also a narrow-angle dual-star facility, designed specifically for studying the massive black-hole at the Galactic Center in the $K$ band (2.2$\mu$m). GRAVITY will allow objects of magnitude up to $K$ $\sim$ 18 to be observed as long as a suitable off-axis reference object is available ($K$ $\sim$ 10) using the 8m UTs. Three spectral resolution modes in the $K$ band will be accessible i.e. $R$ $\sim$ 22, 500 and 4000. The difficulty with this kind of instrumentation is to find a suitable reference target within a region as close as 60\arcsec\ from the source. However, this might not pose much of a difficulty in the case of novae as they are statistically more numerous in the crowded galactic plane.

VLTI/MATISSE (Multi-AperTure mid-Infrared SpectroScopic Experiment) will be able to combine four UT/AT beams of the VLTI simultaneously in the $L$ (3.0-4.1$\mu$m), $M$ (4.3-5$\mu$m) and $N$ (8-13.5$\mu$m) bands. This mid-infrared instrument will include several spectroscopic modes, from  simultaneous observations in the $L, M, N$ bands at $R$ = 30  to observations at $R$ = 1000 in the $L$ band only. This gives access to spectrally resolved observations in the Br$\alpha$ 4.05$\mu$m,  Pf$\epsilon$ 3.04$\mu$m, Pf$\delta$ 3.3$\mu$m and  Pf$\gamma$ 3.74$\mu$m lines  with a velocity resolution of $\sim$300\kms, well suited for the study of nova ejecta.

VLTI/MATISSE and VLTI/GRAVITY will offer the measurement of interferometric observables on six baselines, providing greatly improved imaging capabilities over first generation instruments.
To date, only CHARA/MIRC, CHARA/VEGA and PIONIER/VLTI are able to recombine 4 telescopes and CHARA/MIRC should soon be able to recombine the 6 telescopes of the CHARA array in the near-IR. CHARA/MIRC has provided some of the best interferometric images, encompassing fast rotators \citep{2005ApJ...628..439M}, the interacting binary $\beta$\,Lyrae \citep{2008ApJ...684L..95Z} or the eclipsing long-period binary $\epsilon$\,Aur \citep{2010Natur.464..870K} and this instrument has  great potential for observing novae. With 0.3mas spatial resolution and spectral resolution up to 30000  in the visible (0.5-0.85$\mu$m), CHARA/VEGA is also well suited to study novae \citep{2009A&A...508.1073M}. The medium spectral resolution mode at $R$ = 6000 ($\delta V=60$\kms) offers the best compromise in terms of sensitivity with a limiting magnitude in visible of about 5-6, while the high spectral resolution requires sources brighter than the fourth magnitude\footnote{Moreover, CHARA/VEGA benefits from  efficient fringe tracking in the near-IR with the CHARA/CLIMB instrument.}.

\section{Initial pioneering studies}\label{s:old}
The first published interferometric observations reported on novae are from \citet{1993AJ....106.1118Q}.
Nova Cygni 1992 was observed with the Mk III Interferometer ten days after
maximum light. Combining the Uniform Diameter,  measured to be 5.1 mas
in a 10 nm wide filter centered on the H$\alpha$ line, and associating this measure with an average observed
expansion velocity of about 1100\kms gave a distance to the nova of $\sim$2.5 kpc. This was in
good agreement with later results obtained with the Hubble Space Telescope \citep{1995A&A...299..823P}.

\begin{figure}
\centerline{\includegraphics[width=15cm]{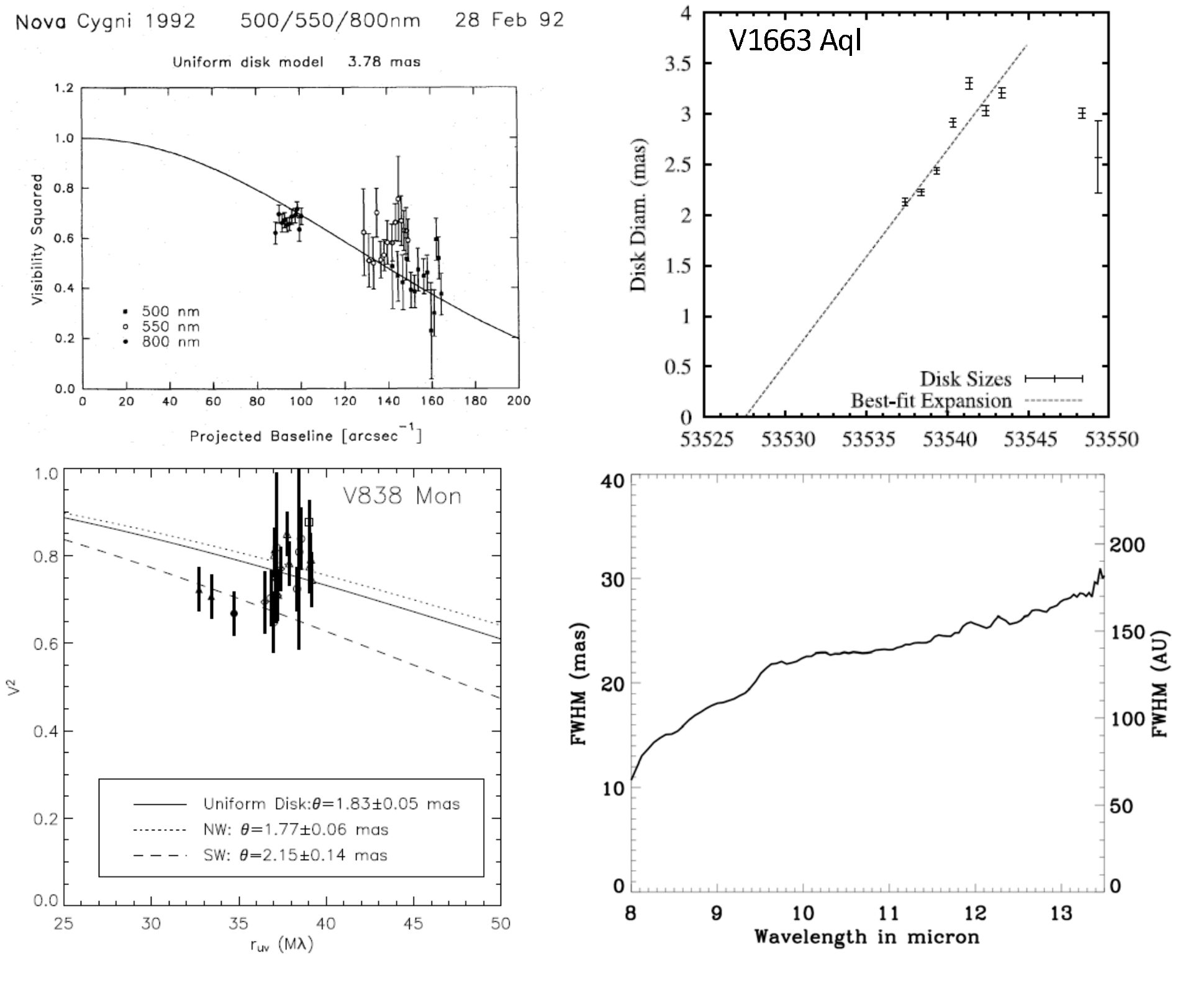}} %
\caption{{\bf Upper left:} Mark III visibilities of the nova cygni 1992 \citep{1993AJ....106.1118Q} {\bf Upper right:} Time variation of the angular extension of the classical nova V1663\,Aql measured by PTI \citep{2007ApJ...669.1150L}{\bf Lower left:} Visibilities of V838\,Mon obtained by PTI \citep{2005ApJ...622L.137L}. A strong dependency on the baseline orientation indicates a flattened source. {\bf Lower right:} First dispersed visibility measurement obtained in 2011-12 of V838 Mon in the mid-IR with the MIDI/VLTI instrument, almost 10\,yr after the event.  \label{f:old}}
\end{figure}

The second successful interferometric observation of an explosive nova-like event was the observations with the PTI of V838\,Mon \citep{2005ApJ...622L.137L} that was not a true nova eruption but rather a rare and intriguing phenomenon best interpreted as a merger event (see discussion Sect.7). The publication of the next interferometric observation of a true nova had to wait until 2007, when Lane et al. (2007) resolved the classical
nova V1663\,Aql (2005) using the PTI observing it over a period between 5 to 18 days after maximum light. The time for the brightness to decline by 2 magnitudes ($t_2$) was about 16 days for the object, classifying V1663\,Aql as a 'fast nova'. They directly measured the shape and size of
the fireball which was found to be asymmetric.
From the apparent expansion rate, estimated to be  0.21$\pm$0.03 mas/day, a distance to the source of 8.9$\pm$3.6 kpc was inferred. Assuming a linear
expansion model, the start of the eruption was approximately estimated to be  4 days prior to the time when peak brightness was reached. The period 2006-2007 was a defining moment in the interferometric studies of novae because   many studies were undertaken during this time arising from the eruption of the  famous recurrent nova RS Oph  in  February 2006.

\section{The recurrent nova RS Oph}\label{s:rsoph}

RS\,Oph is a famous recurrent nova, that has had recorded outbursts in 1898, 1933, 1958,1967, 1985 and 2006. The central system comprises of a high mass WD in a 455 day orbit with a red giant (M2III) companion. The outburst of  RS Oph on February 12, 2006 was observed with four arrays worldwide, and among them for the first time with the VLTI, representing a unique event in the short history of optical interferometry. The interferometric observations were insufficient to allow any image reconstruction to be made that could be compared with the radio-interferometry and HST images of the object that showed  an equatorial ring \citep{2006Natur.442..279O} and a full bipolar nebula \citep{2007ApJ...665L..63B}, respectively.

\begin{figure}[h!]
\centerline{\includegraphics[width=15cm]{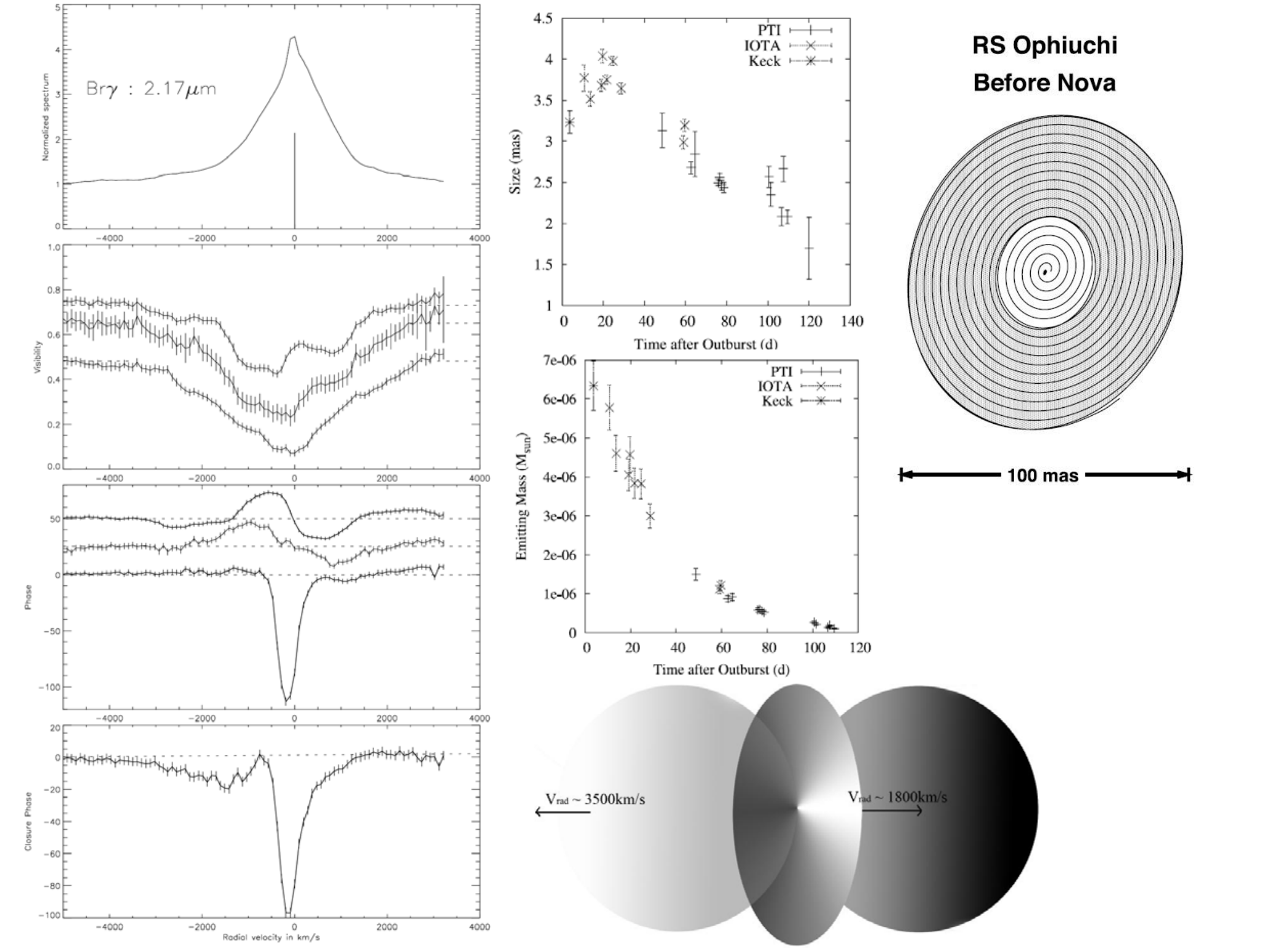}} %
\caption{{\bf Left:} AMBER/VLTI dispersed visibilities of the nova RS\,Oph \citep{2007A&A...464..119C} obtained 5.5 d after the outburst (from top to bottom: spectrum, differential visibilities and phase, phase closure).   {\bf Lower middle:} Visibilities of V838\,Mon obtained by PTI \citep{2005ApJ...622L.137L}. {\bf Middle} Angular diameter measurements from PTI, IOTA and Keck instruments \citep{2007ApJ...658..520L, 2006ApJ...647L.127M}, and below the estimate of the emitting mass using a simple model of an optically thin free-free emitting sphere. {\bf Right:} Schematic model of the low potential emission detected by the Keck nulling experiment before the blast wave reached these regions \citep{2008ApJ...677.1253B}. \label{f:rsoph}}
\end{figure}

{\bf First IOTA/Keck/PTI observations}:
\citep{2006ApJ...647L.127M} observed the erupting source using the IOTA (H band), Keck and early PTI interferometers (K band) at multiple epochs. They concluded that
the characteristic size of $\sim$3 mas hardly changed over
the first 60 days of the outburst, dividing the dataset into three main parts, days 4-11, 14-29 and 49-65. The insufficient time sampling, baseline coverage and the level of uncertainties hampered the detection of a temporal evolution of the angular diameter of the source. Importantly, the emission was
also found to be significantly asymmetric, evidenced
by nonzero closure phases detected by IOTA. Some interesting ideas were proposed in this study, particularly the suggestion that pre-existing material was being seen outside the binary system.

{\bf PTI observations}:
Using
the PTI interferometer, \citet{2007ApJ...658..520L} performed
observations of RS Oph, resolving the emission from
the nova for several weeks after the outburst. The dataset, more extended and more consistent that the one reported in \citet{2006ApJ...647L.127M} provided a clear detection that the near-IR source initially expanded to a size of about 5 mas, began to shrink around
day 10, and reached $\sim$2 mas by day 100 (Fig.\ref{f:rsoph}). These important results are extensively discussed below.

{\bf VLTI observations}:
Following the submission of a DDT proposal which was rapidly accepted, 2h of AMBER/VLTI observations were obtained at $t$=5.5d (RS\,Oph was observable only at the end of the night). The outcome of this limited set of observations, are described in \citet{2007A&A...464..119C} containing a spectrum, three dispersed visibilities and phase, and a dispersed closure phase (Fig.\ref{f:rsoph}). The angular diameters in the $K$ band continuum, and in two lines, Br$\gamma$ 2.1655$\mu$m and HeI 2.058$\mu$m were measured. The $K$ band continuum, dominated by free-free emission, had at this date an angular diameter of 3 x 2 mas (assuming a Gaussian distribution of the flux) which is larger than the  system dimensions (semi-major axis $a$ = 0.7mas  assuming $D$ = 1.6kpc). The Br$\gamma$ emission line was more extended (5 x 3 mas), originating clearly from the ejecta, and the HeI emission is even more extended than the Br$\gamma$ one (6 x 4 mas) here. This latter line must be formed in a region very close to the shock caused by the fast ejecta colliding with the slow red giant wind.

The spectral dispersion of AMBER ($R$ = 1500, $\delta v \sim 200$km.s$^{-1}$) provided information on the kinematics in the Br$\gamma$ line. In particular the phase of the fringes can be, to the first order, associated with the photocenter of the emission through the line and thus provide, through an analysis of the Doppler effect, some kinematical information on the source. The phases were clearly split between a 'low velocity' signal in the bulk of the line and a 'large velocity' inverted signal. The low velocity signal was interpreted in the frame of an expanding structure, tracing primarily the ejecta in the equatorial plane. The interpretational outcome of these measurements was that the bright half ring seen in the radio images \citep{2006Natur.442..279O} appears to be  the {\it rear} side and not the front side of a tilted equatorial structure. The large velocity signal was best observed in the East-West baselines, and was attributed to the jet-like feature observed in the center of the bipolar lobes (Fig.\ref{f:rsoph}).

{\bf The Keck Nulling experiment}: \citet{2008ApJ...677.1253B} reported
$N$ band observations of RS\,Oph using
the Keck Interferometer 3.8 days after the outburst. Their data showed evidence of enhanced neutral
atomic hydrogen emission and atomic metals including silicon located in the inner spatial region near the
white dwarf. They also reported presence of nebular emission lines
and evidence of hot silicate dust in the outer spatial
region, centered at $\sim$17 AU from the white dwarf,
which were not found in the inner regime (Fig.\ref{f:rsoph}). Their
results support a model in which the dust appears
to be present between outbursts. The key question is to know whether the circumbinary material has a definite impact on the shaping of the ejecta or not.

\section{The dusty classical nova V1280 Scorpii}\label{s:v1280}
\begin{figure}[h!]
\centerline{\includegraphics[width=15cm]{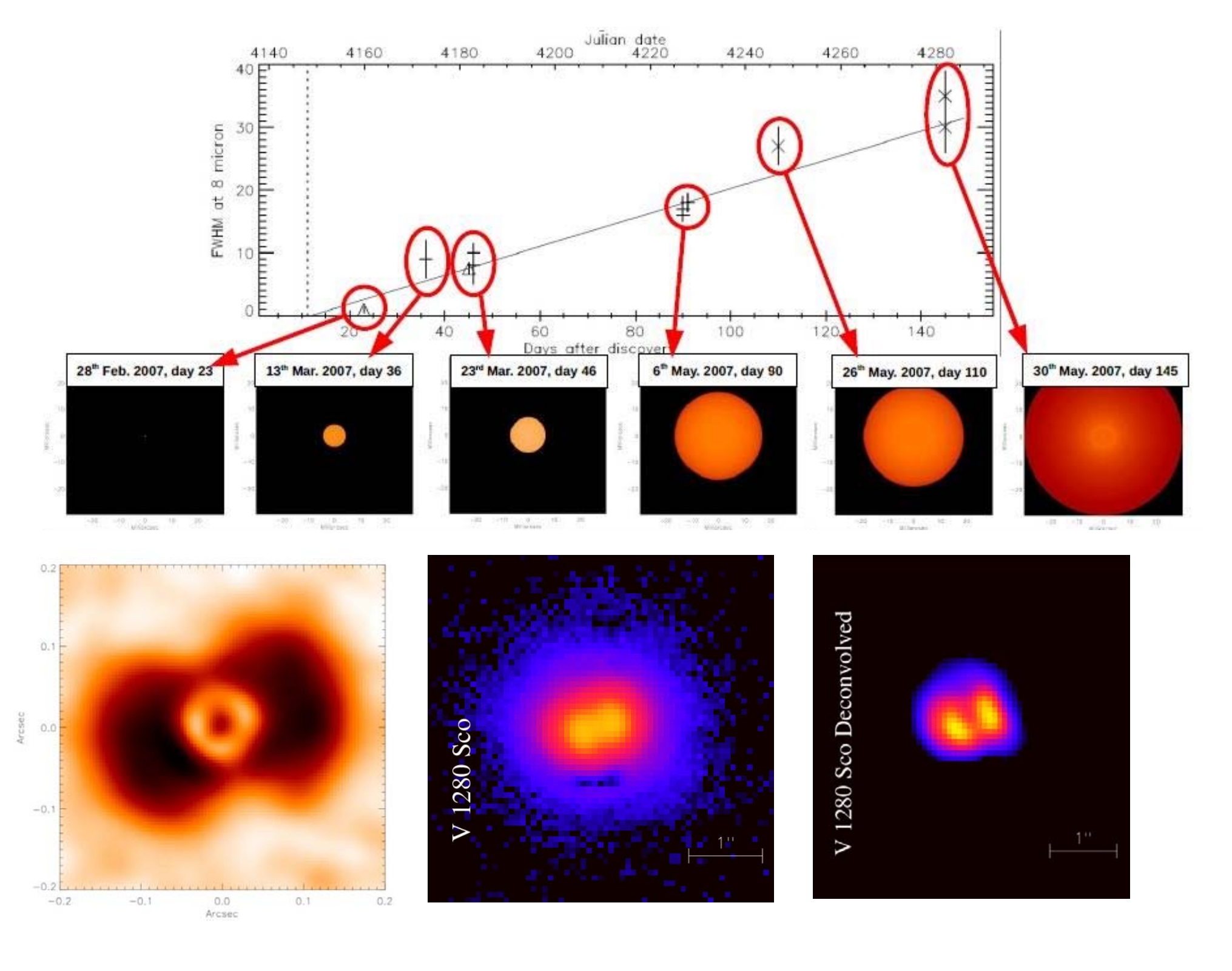}}
\caption{{\bf Up} Temporal evolution (first 150 days, in 2007) of the angular diameter of the dust shell around V1280\,Sco as estimated from a Gaussian model applied to the MIDI/VLTI and AMBER/VLTI measurements. {\bf Bottom left} NACO/VLT observations of the source obtained in 2010 in the K band (the central unresolved source is removed and the colors are shown in inverse)with the dust emission showing a  striking bipolar shape. {\bf bottom middle and right} VISIR/VLT observations obtained in 2011 at 8.59$\mu$m. The emission originates clearly from the polar caps rather than the equatorial regions.  \label{f:v1280}}

\end{figure}

Approximately a year after the RS\,Oph outburst, another bright nova outburst took place in  V1280\,Sco which reached a peak brightness of 3.8 in the $V$ band. AMBER/VLTI observations with the 1.8m ATs were requested, but within a few weeks after discovery the nova began to fade sharply  in the visible due to the onset of dust formation. Most of the planned near-IR AMBER observations were hence shifted to  MIDI in the mid-IR while retaining the plan to use the ATs. At $t$ = 23 days, V1280 Sco exhibited a compact core of $K$ band emission ($\leq$3mas), dominated by an optically thick free-free photosphere (the extended ejecta are quasi-transparent), while at $t$ = 36 days, the angular diameter reached suddenly $\sim$13mas, when the dust formed at high rate, dominating the infrared flux. While the $N$ band flux kept steadily increasing from March 2007 to June 2007, the $V$ magnitude continuously faded. As a consequence the UTs had to be requested for the last set of observations  as the source could not be tracked any longer with the ATs. Owing to the very limited $uv$ coverage at each epoch, radiative transfer models were computed using the 1D DUSTY code to model and explain the observed data. These observations provided determination of the apparent linear expansion rate for the dust shell of 0.35 $\pm$ 0.03 mas per day, and the approximate ejection time of the matter in which the dust formed (t$_{ejec}$ = 10.5$\pm$ 7 d), i.e. close to the maximum brightness (Fig.\ref{f:v1280}).

\citet{2006ApJ...653L.141D}, in their studies of V1280 Sco, proposed  generic conditions favorable for dust formation to take place in novae. Their  spectroscopic studies and modeling showed that
the presence of lines of low ionization species such as Na and Mg in the early IR spectra, as seen in V1280\,Sco, were  a trait of low temperature conditions and conducive for  dust formation. Sustained mass loss was also suggested from following the P-Cygni profile evolution of the Brackett $\gamma$ line. Line profiles, giving velocity information,  were published in
\citet{2008MNRAS.391.1874D} in the near-IR domain, and by \citet{2010PASJ...62L...5S}
in the visible. \citet{2010PASJ...62L...5S} reported multiple high-velocity narrow components
in the NaID line from 650 to 900 km.s$^{-1}$. They interpreted the formation of these clumps
as a consequence of a strong shock between the slow, dust forming ejecta, and a
fast wind ($\sim$ 2000 km.s$^{-1}$) generated during the second brightening of the
source that took place about 110d after discovery \citep{2008A&A...487..223C}. Surprisingly, P-Cygni lines were systematically observed even 4yrs after the outburst, indicative of a sustained and significant mass-loss rate \citep{2011ApJS..197...31S}. Using a representative value of the velocity of 500 km.s$^{-1}$ as suggested by the P-Cygni profiles and other kinematic information,  \citet{2008A&A...487..223C}
inferred a distance of 1.6$\pm$0.4kpc. However, a recent estimate of the distance of
630$\pm$100\,pc has been derived from high cadence space-based light-curves \citep{2010ApJ...724..480H} that cover the first 20 days after discovery.

Based on the measurements of the expansion of the dusty nebula inferred
from the initial VLTI measurements, it was expected that the ejecta  of V1280 Sco
could reach a  spatially resolvable size within a few years. It was thus expected that high resolution instruments mounted on large aperture telescopes such as the 8.2m UTs at the VLT could resolve such a shell. Follow-up observations were obtained in 2009, 2010 and 2011 by the NACO and VISIR instruments attached to the VLT. The first observations, secured in summer 2009, more than two years after the outburst, revealed a striking bipolar nebula which is seen to be expanding with time (Chesneau et al., submitted). Fortunately, most of the projected baselines of the VLTI observations were oriented roughly (i.e. within 20\deg) in the direction of the major axis, so that the expansion measurements are still consistent and meaningful for the distance estimate. Yet, the association of the expansion measured in the sky in the direction of the major axis by Doppler velocities measured into the line of sight is not obvious anymore, and the distance estimate subject to potential large biases.

Coming back to the VLTI data, some hints of significant departures of symmetry were found in the latest dataset, secured 110 days after outburst, indicating that the shaping process took place early. However, these observations did not provide enough information to identify clearly the shaping process.

\section{The recurrent nova T Pyx}\label{s:tpyx}

The most recent eruption of recurrent nova T Pyx took place on 14 April 2011, its sixth known eruption. Interferometric observations of this outburst were made in the near-infrared by \citet{2011A&A...534L..11C} on days between 2.4 to 48.2 days after the outburst with the CLASSIC
recombiner located at the CHARA array and with
the PIONIER and AMBER recombiners located at
the VLTI array. The $H$ and
$K$ band continua were observed to be slowly expanding (i.e. $\leq$300\kms), and with a spherical appearance. The Br$\gamma$ emission line also exhibited a slow expansion in the plane of the sky (i.e. $\leq$700\kms), in contrast to the information provided by the Doppler width of the line. A complex but structured phase signal was also observed that led the authors to conclude that the
outburst could be most simply interpreted by a model
where the ejecta material has a  bipolar shape oriented nearly face-on. Such a low inclination (i=10\deg) was already inferred from crucial spectroscopic observations from \citet{2010MNRAS.409..237U}.

The material in the shell was assumed to follow a homologous flow as generally expected in bipolar shaped structures (for e.g in several planetary nebulae). A semi-quantitative mode was used to interpret these data. Given the limited amount of data, the model worked surprisingly well in reproducing the double S-shaped phase  signal that is strong evidence of a bipolar velocity field.

These findings have some important implications for the interpretation of the slowly expanding ejecta observed by the HST \citep{2010ApJ...708..381S, 1997AJ....114..258S}. In the semi-quantitative face-on model, the densest regions that are located close to the plane of the sky exhibit radial velocities limited to 600\kms, while the fast flows are focused in the direction of the line of sight, and diluting probably rapidly. Some caution must therefore be taken about the potentially simplistic affirmation that the 1866 eruption was a slow eruption with an isotropic ejection velocity of 500-715\kms, and the equivalently potentially simplistic hypothesis that each subsequent recurrent nova eruption sent a fast, low mass spherical shell \citep{2010ApJ...708..381S}. The bipolar model complicates this picture. The HST observations and the study of the old nebula is a complex task and a detailed (spatially resolved) Doppler and polarimetric analysis of a set of individual clumps is needed to clarify this issue.

\section{Other sources}\label{s:intro}

{\bf Symbiotic novae}:
Symbiotic novae are observed in well separated binary systems involving a WD and a high mass-losing giant star. The eruption is observed as a very slow nova-like outburst of 9 to 11  magnitudes in amplitude.
For optical interferometers, classical nova events in symbiotic systems  offer several advantages: the time scale is very long and one can study the expanding ejecta {\it and} potentially resolve the two stellar sources. Symbiotic systems are well separated, and the separation, when the period is in years is of the order of 1\,mas. Yet, one must keep in mind that separating the components is possible only when the WD is in eruption, and the two fluxes are reasonably equivalent (within a flux ratio range of about 1/10-1/50). Most of the interferometers currently observe in the infrared where the flux of the cool giant overwhelmingly  dominates over the flux from the WD and the accretion disk such that these observations are challenging during quiescence.

The number of symbiotic systems that have experienced an observable outburst is limited at most to  about a dozen sources. Among these, the system HM\,Sagittae was observed by the VLTI \citep{2009A&A...493.1043S, 2007A&A...465..469S}. The data showed that the dense dusty wind of the Mira star was only weakly perturbed by the hot WD. The nova outburst of V407 Cyg in 2010 was unfortunately not observed by any optical interferometer, being even at maximum too close to the sensitivity limits. It's close resemblance to the well known symbiotic-like recurrent nova RS Oph \citep{2011MNRAS.410L..52M} and the bipolarity of the ejecta makes this an attractive source to study how the shaping of the ejecta occurs \citep{2012MNRAS.419.2329O}.

{\bf V458 Vul}:
Nova Vulpeculae 2007 No. 1 (alternatively designated as V458\,Vul) was a very interesting source. It was a fast nova with t$_2$=7d and t$_3$=15d  respectively. The nova exploded in the center of a PN which is very rare suggesting that the 'nova' stage is very young \citep{2008ApJ...688L..21W} following a recent post-common envelope phase \citep{2010MNRAS.407L..21R}.
Moreover, the light curve exhibited some brightening episodes that were discussed in \citet{2011PASJ...63..159T}. The nova was spatially resolved by the PTI very early in the $K$ band \citep{2012arXiv1206.5172R}. These observations provided a distance estimate of 10-11\,kpc which was of importance for estimating the luminosity of the central source and the surrounding Planetary Nebula. This source should be scrutinized extensively in the following years.

{\bf V5583 Sgr}:
Nova Sagittarii 2009 No. 3 (alternatively designated as V5583\,Sgr) was a fast nova  with decline times of  t$_2$=4.5 and t$_3$=8.8 days respectively \citep{2010JAVSO..38..193K, 2010CBET.2265....5M}. Three observations with VLTI/AMBER were attempted between 6 to 8 days after the discovery, at one day intervals, in medium resolution mode. The spectra exhibited strong emission lines such as Br$\gamma$ and Br8 1.95$\mu$m, that appeared to be flat-topped and rapidly raising in intensity.  Gaussian fits to the Br$\gamma$ line profiles provided  FWHM values of 2817\,km.s$^{-1}$, 2940\,km.s$^{-1}$ and 3061\,km.s$^{-1}$ for days 5, 6 and 7, respectively. The Full Width at Zero Intensity (FWZI) increased from 3200\,km.s$^{-1}$ to 3500\,km.s$^{-1}$ ( error $\sim$  300\,km.s$^{-1}$)) between these three days. Structures were seen on the top of the profiles of
the  broad lines.

With such a large expansion velocity, the ejecta could be expected to rapidly attain a spatially resolvable size provided it is nearby. For e.g. assuming a velocity for the ejecta of 2500 km.s$^{-1}$ or more, the source is larger than 2\,mas in 6 days at a distance of 4kpc (note: the distance to the object was uncertain at that stage but the above arguments provided sufficient motivation to attempt observing it). It may be noted that an Uniform Disk with an angular diameter of 2mas can be significantly resolved with a baseline of 100m in the $H$ and $K$ band, considering an error bar on the measurements of 15\%. However, in our observations, the nova remained unresolved by the interferometer providing a lower limit for the distance of about 7\,kpc. \citet{2010JAVSO..38..193K} used the MMRD relation to infer a distance of 11$\pm$3kpc, putting V5583\,Sgr on the other side of the Galaxy, in the Bulge at 1.1-1.4 kpc above the Galactic plane.

{\bf V838 Mon}
Though not a nova formally, the outburst of this object  shared many aspects of the nova phenomenon.  V838\,Monocerotis is an eruptive variable which gained an iconic status in 2002 when it
brightened by 9 magnitudes in a series of outbursts (reaching V$\sim$6.8), and eventually developed a
spectacular light echo both in the optical \citep{2002A&A...389L..51M, 2003Natur.422..405B} and infrared \citep{2006ApJ...644L..57B}. The eruption was unlike classical novae in that the effective temperature of the object dropped and the spectral type evolved into that of a very late M to L type star. A promising scenario for the origin of this event is the merger of two stars \citep{2003ApJ...582L.105S} though a mechanism of a planetary capture by an expanding red giant cannot be ruled out \citep{2003MNRAS.345L..25R, 2004ApJ...615L..53B}.

PTI observations  performed in Nov - Dec 2004, about 34 months after the event, were able to resolve the source and to provide the first direct measurement of its angular size \citep{2005ApJ...622L.137L}. Assuming a uniform disk model for the emission, the above authors derived an apparent angular diameter of 1.83$\pm$0.06 mas. For a nominal distance of 8$\pm$2 kpc, this implied a linear radius of 1570$\pm$400\rsun. However, the data were better fit using an elliptical model with a major axis of 3.57 x 0.07 mas  (i.e. the minor axis being unresolved) or a binary at Position Angle (P.A.) 15-36\deg. Polarimetric observations performed during the outburst \citep{2004A&A...414..591D, 2003ApJ...598L..43W} also indicated a flattened source. The intrinsic polarization angle after its second outburst on 2002 February 8 was 127\deg. Right from the outburst beginning in 2002 till present day measurements,  the object's SED and other modeling suggests it has a central source with a temperature broadly in the range of 2200 to 2500K.

The merger scenario proposed to explain the outburst of V838 Mon has two strong predictions for the properties of the merger remnant. First, a disk or disk-like structure should form around the central object. It would contain the excess  angular momentum, which could not be stored in the star and which originates from the accreted body. From the same reason the star is expected to shrink with time. Another prediction is that the star would be expected to be a fast rotator and exhibit a large magnetic activity \citep{2007MNRAS.375..909S}.
Some of these predictions, the presence of a disk in particular,  are testable interferometrically.

Recently, MIDI/VLTI observations of the source were performed. Despite a limited $uv$ coverage, the source seems to exhibit a flattening. The P.A. of the major axis of this flattened structure is in close agreement with the early polarimetric data. Awaiting further complementary data, investigations similar to those performed for Sakurai's object for instance \citep{2009A&A...493L..17C} are planned to be carried out in order to derive the basic parameters of the dusty circumstellar environment of V838 Mon.

\section{Interferometric observables and constraints on physical processes}

\subsection{Interferometric observables}
Until now, most of the interferometric observations were performed recombining the light from two or three telescopes only. This does not give enough information to perform an image reconstruction process and understanding the observations relies on a  on semi-quantitative interpretation based on a few observables. The visibility is proportional
to the Fourier transform of $I_{\lambda}(y,z)$ of the sky-projected spatial distribution of the source $I_{\lambda}(y,z)$, hereafter called "intensity map". The absolute visibility (modulus of the complex visibility) mainly depends on the equivalent size
of the source in the direction of the baseline. A fit of absolute visibilities
as a function of baseline length allows us to estimate the
equivalent radial intensity profile. Since a few visibility measurements are obtained, a set of simple models are frequently used. The Uniform Disk is a model of a circular source of uniform brightness. Theoretically, one visibility measurement is enough to constrain the diameter of the source provided that one knows in advance that the source can be ascribed by such a model. This model is mostly applicable in measuring the diameter of a normal star (neglecting the effect of limb-darkening).
For  diffuse sources, such as dusty objects, a Gaussian model is preferred.
The visibility for an Uniform Disk of diameter $\Theta$ is given by the following expression:

{\large $V_{UD}(f) = \frac{J_1(\pi\Theta f)}{\pi\Theta f}$}

where $f$=$B$/$\lambda$ is the spatial frequency of the projected baseline $B$ at the wavelength $\lambda$, and $J_1$ is the first order Bessel function. The value
for which the visibility becomes zero is $B$ = 1.22$\lambda$/$\Theta$. Measuring the visibility null is the most accurate way to infer $\Theta$, but measurements made with shorter baselines if accurate enough can also provide good estimates.

The visibility of a Gaussian
 distribution can be calculated as:\\
 {\large  $V_{Gauss}(f) = V_0~exp(-3.56\,f^2\,\Theta^2)$}\\ where $\Theta$ is the FWHM (measured in arc-seconds) of the Gaussian distribution and $f$ the spatial frequency in units of arcsec$^{-1}$.

An erupting source can be represented by an optically thick core, best ascribed by a Uniform Disk (UD) and an  envelope described by a Gaussian. The resulting visibility is a function of the core UD diameter, the Gaussian FWHM and the flux ratio $F_r$ between the two sources.

 {\large  $V_{obs}=F_r*V_{UD}+(1-Fr)*V_{Gauss}$}

To constrain these parameters at least three visibility measurements, optimally at very different spatial frequencies i.e. projected baselines,  are necessary at a time, assuming that the source is spherical.

\subsection{Probing the P Cygni phase}

A nova outburst is triggered by a thermo-nuclear runaway reaction in the accreted envelope of matter on the WD's surface. In the rising phase the envelope expands with increasing luminosity. During this phase, the expanding sphere may well be described by a Uniform Disk. The source may be approximated at this stage in a relatively simple manner: an optically thick photosphere with little material outside it. For distance estimation and a better knowledge of the fireball stage, it is important to resolve the outbursting source at the earliest moments. This is best performed in the visible using the longest baselines of the CHARA array that provide a spatial resolution of the order of 0.2\,mas at 0.6$\mu$m. It is important also to study the formation of P Cygni profiles. Prominent P Cygni profiles in  a nova outburst are inevitably seen
and reported at maximum light and for a few days following it and this phenomenon is expected to arise from regions close to the WD expanding photosphere on the scale of sub-mas \citep*{2011MNRAS.415.3455R} when the novae spectrum is around maximum light. As mentioned earlier it is expected from theoretical considerations of the photospheric optical depth
that the photosphere radius $r_{ph}$ $\leq$ 100 $R_{sun}$ at maximum light (Kato \& Hachisu (1994);  Hachisu \& Kato, 2001;  Hachisu \& Kato, 2006), with an angle of 0.2-1\,mas at 1-5 kpc.

 Interferometry can currently measure visibilities at different wavelengths across a line profile at suitable spectral resolutions (R=6000 with VEGA/CHARA from 0.55 to 0.85$\mu$m), and some phase information can differentially be retrieved by comparing the fringes in the continuum (to which a reference phase of zero is attributed) to the fringes in the lines. This spectrally dispersed phase information can in first order be associated to a change of the photocenter position of the source in each spectral channel by comparison to the continuum. When coupled with modeling, one can in principle determine where the emission associated with each spectral element arises from and thus infer some kinematical information. This technique was used with the AMBER/VLTI instrument to infer the double structure of the velocity field in RS\,Oph and T\,Pyx, when the emission lines were almost fully developed \citep{2011A&A...534L..11C,2007A&A...464..119C}.

The generic formation of P Cygni profiles can be  understood by considering
a spherical symmetric outflowing wind in which the velocity
necessarily increases outwards i.e., the wind is accelerated outwards till it reaches a terminal velocity (Lamers \& Cassinelli 1999). To the outside observer it
is the matter in the form of a tube in front of the stellar disc which scatters light from the continuum of this star that  is
responsible for the absorption component of the P Cygni profile.
The ratio of the strength of the emission and absorption components of the P Cygni profile depends on the size $r_w$ of the wind region
(i.e., size of the ejected material) relative to the size of the optically thick part of the emitting region. When the wind region is large compared to the source's size we expect emission
to dominate - this can be seen geometrically as the volume of the emitting gas becomes much larger than the volume causing the absorption component.
 At such an epoch, it is therefore reasonable to use an
approximation, to the first order, that the wind region size $r_w$ is of the order of the stellar size $r_{ph}$. Thus from the spectrally dispersed visibility curve across a P Cygni profile, one should be able to estimate the sizes of different regions viz the regions causing the absorption and the emission.  As the ejected matter (the wind region) keeps expanding to larger sizes at
 later times following the maximum, the emission component strengthens and finally begins to dominate. The spectrally dispersed visibilities should be able to track this evolution, and the dispersed phases provide information on the wind acceleration using the Doppler information and potentially detect any hint of asymmetries that may occur as early in the nova development.

\subsection{Novae distances}

Optical interferometry has considerable potential in measuring  the distance to a close-by nova with  accuracy if the spatial geometry  of the source is known and not too complex. The restriction to nearby sources arises only from brightness considerations given the  current detection  limits of the instruments involved. Resolving the milliarcsecond-scale emission and measuring the temporal change in angular size of the emission zone provides information on the angular parallax. When this is combined with velocity information of the emitting regions, obtained spectroscopically from line profiles (or better by the interferometer itself), a geometric distance estimate (and hence luminosity estimate also) can be obtained.

Although the approach  may appear straightforward, there are caveats. The distance estimates compare the observed angular size of a nova shell with its linear size (the so called Uniform Diameter) as inferred from the optical interferometry measurements.
But certain precautions must be taken to ensure that the result is accurate:\newline
- the measurements should be made as early as possible to be in the fireball stage, when the optically thick envelope is best ascribed by a uniform disk,\newline
- Spectroscopic observations are needed quasi-simultaneously to provide kinematic information on the velocity.\newline
- Apply a projection factor relating the Doppler expansion velocity to the true one. The determination of the projection factor appear to be a key point for improving the distance determination in Cepheids \citep{2004A&A...428..131N, 2005A&A...438L...9M}. It must be borne in mind that even if a spherical shell is expanding with an uniform velocity, the observed line profile will be considerably broadened because of several factors. There will be large  line-of-sight Doppler broadening because the source is unresolved in the spectrograph slit and the line of sight takes in all velocity components from the expanding source arising from the cos($\theta$) dependence. Additionally there will be smaller contributions from turbulent broadening and thermal broadening (which will be marginal at the typical 10,000K temperature expected in nova ejecta). Thus some amount of modeling of the line profiles may be needed to infer what the "true" velocity is. \newline
- The approach is valid for dusty nova as long as the source is optically thick at the wavelengths of observations,
- The approach will be more reliable when the geometry of the source is known to be spherical; deviations from spherical symmetry will give rise to complications.

\subsection{The wind ejection scenario}

As  mass-loss driven by a radiative wind sets in,  the ejecta are launched at velocities much larger than the expansion velocity of the photosphere during the fireball phase. At this stage, the ejecta emits  copious amounts of free-free emission whose contribution to the infrared continuum cannot be neglected. As seen from an optical interferometer the source is now composed of two components viz the first is a stellar component where the photosphere initially expands and then rapidly shrinks to the point where it can be considered as unresolved by the instrument, and the second component being the expanding optically thin envelope  whose appearance can be, to the first order, approximated  by a Gaussian.

\citet{2006ApJS..167...59H} formulated a possible universal decline law for classical nova, based on the application of the optically thick wind model \citep{1994ApJ...437..802K}. The model is particularly applicable during the declining phase, and \citet{2005ApJ...631.1094H} had to include free-free emission from an optically thin plasma outside the photosphere to explain the power-law declines observed in the near-IR lightcurves of selective novae. We note that a library of models was published by \citet{1997ApJS..113..121K} and is available online\footnote{http://vizier.cfa.harvard.edu/viz-bin/VizieR?-source=J/ApJS/113/121}. In this context we caution against the systematic use of a Uniform Disk as being a sufficient condition,  while modeling interferometric data,  to account for the complex appearance of the source. At any time, the visibility of the source is likely influenced by  two contributing factors viz. one component arising from the shrinking optically thick photosphere and the other from the  expanding optically thin ejecta. Moreover, the contribution from these components is highly wavelength dependent, since the free-free flux rises rapidly with wavelength. The difficulty in practice is that only few interferometric measurements are performed at a time, preventing the fitting of the data by models involving more than one parameter. \citet{2007ApJ...658..520L} estimated the mass of the ejecta from the time-dependent Gaussian FWHM measurements and an optically thin model of the free-free emission of the source that neglects the contribution from any compact optically thick source. They used  the $H$ and $K$ light curves to estimate the temperature, and the Gaussian extensions\footnote{This information included the minor/major axes dimension differences that determined an ellipsoid of revolution for the emitting region.} to constrain the source size and volume at different epochs in  time.
The assumptions of the model are that the emission is optically thin, and the electron density is uniform within the volume. In their Figure 4, one can see that they get a consistent estimate of the emitting mass ranging from $\sim$6 x $10^{-6}$\msun at earlier times to less than $10^{-6}$\msun 110 days after the outburst. A possible issue with this model is that it neglects any emission from an optically thick core. This issue is more serious at early times. As time goes by, the central source shrinks to the point it is so small that its flux is negligible compared to the free-free emission from the ejecta. The time when this assumption can be made may be evaluated using a careful comparison between the visible and near-IR fluxes. In the visible, the free-free emission is weak and the black-body emission from the compact erupting source dominates (although scattering emission can also contribute significantly). Ideally, quasi-simultaneous visible/near-IR observations may allow one to disentangle the  individual contributions from the core and that from the expanding ejecta. CHARA is ideally suited for such a  task. The observations of T\,Pyx by PIONIER (at $t$ = 48d) provided the first snapshot multi-baseline data of an erupting nova \citep[Fig.3]{2011A&A...534L..11C}. Most of the interferometric results are based on datasets limited to one or two baseline measurements per day, while a single PIONIER observations provides data from 6 baselines simultaneously. Although limited in spatial frequency range, this dataset illustrates the need of determining at each time the full visibility profile due to the complexity of the source's flux distribution.

\subsection{Dust Formation processes}
When proper physical conditions are met, novae can form dust efficiently at a vigorous rate that apparently
belies their hot underlying central sources \citep{1988ARA&A..26..377G}. The dust forms with various compositions  including silicates, amorphous carbon, silicon carbide, and hydrocarbons. The main reason proposed to explain such a diversity is that CO formation in novae may not reach saturation at later stages \citep{2004MNRAS.347.1294P} so that the C:O ratio does not restrict anymore the types of dust species that can be formed \citep{2005MNRAS.360.1483E}.

Optical interferometry is mature to address the issue of  dust formation. The dust, typically found in the temperature range 1000$\pm$500K, is best studied in the mid-IR. The pioneering study of the dust-forming nova V1280 Sco  by \citet{2007A&A...464..119C} illustrated the potential of the technique to infer the dust formation rate. Yet, these observations were very limited in terms of instantaneous $uv$ coverage with the 2T recombiner MIDI. The VLTI shall soon cover the near-IR to the mid-IR including the intermediate $L$ and $M$ bands with the second generation 4T-recombiner MATISSE instrument. This instrument will also provide some measure of spectral resolution needed to spectrally analyze the dust features encountered. We note however that  whether the dust is clumpy or uniform cannot be answered efficiently for lack of resolution and $uv$ coverage so that a model invoking only a smooth/uniform distribution shall have to be used to interpret the data.

\subsection{The rebrightening phenomenon and the transition phase}
The light curves of classical novae usually comprise of a rapid premaximum rise
and a slower decline from the maximum. The fast (rapidly fading) novae tend to show a relatively smooth
decline, while slow novae tend to show more complex behavior. About 20\% of the novae show jitters/flares or even mysterious quasi-sinusoidal oscillations superposed on the decline  \citep{2010AJ....140...34S}.
An interpretation that has been proposed to explain some rebrightenings is a shock resulting from secondary ejections and their breakout in the optically thick nova winds. By accurately measuring the angular extension of the inner, optically thick core, an interferometer can provide tight constraints on the underlying phenomenon. For instance, the interferometric measurements can help to disentangle between an obscuration event occurring close to the central, compact source from a temporary
expansion/shrinking of the pseudo-photosphere of the nova. In the two component model presented in the previous section, the scheme consists in separating a flux variation due to the compact source or the extended one. Observing the differential impact of a rebrightening in the continuum and in some emission line can also provide further indications about the development of shocks in the vicinity of the atmosphere \citep{2009arXiv0904.2228K}.

The difficulty of observing such phenomena is that they are usually observed super-imposed to the decline of the nova so that the sensitivity limits of the interferometers are rapidly reached.

\subsection{Shaping bipolar morphologies}

In this final section, we deal with the bipolar shaping of novae ejecta, a shape which if not ubiquitous is certainly being encountered rather frequently in recent spatially resolved images of novae. The first outcome of these interferometric studies is the observation that remnant shaping occurs very early in a nova outburst. This finding makes the analysis described above considerably more complex than in the case of a spherical geometry. For instance \citet{2012AAS...21932003B} re-analysed  the photometric observations (including visible) of the dust forming nova V868\,Cen using the DIRTY code and showed that while a spherical shell model could fit the data of individual nights, it could not match the temporal evolution of the nova. More complex ellipsoidal geometries are required  with equatorial (torus model), tropical, and polar overdensities.

{\it Circumbinary reservoir} An important aspect  that has been given consideration recently is whether  a circumbinary reservoir of material can affect the ejecta and to what extent. High resolution post-outburst spectra of novae also suggest  episodes of enhanced mass transfer originating from the secondary star that result in the formation of discrete components of circumbinary gas \citep{2011arXiv1108.4917W}. The presence of such matter manifests itself most strikingly in the case of recurrent novae with a red giant secondary which have a substantial mass-loss rate. Particularly discussed in this context is the case of RS\,Oph and its most recent outburst in 2006. A reservoir of mass was  proposed very early by \citet{2006ApJ...647L.127M}, which was further confirmed by the detection by the Keck Nuller of material excited by the flash {\it before} being swept by the blast wave \citep{2008ApJ...677.1253B}, and by subsequent mid-IR observations \citep{2008ASPC..401..260W}. The authors proposed a model in which material appears to be present between outbursts, lying preferentially in the orbital plane. This model is reminiscent of the notable study of \citet{1999ApJ...523..357M}. The question is whether or not this material can affect the flow of the nova ejecta. Consistent modeling,  including the accretion to the  outburst stage, was performed by \citet{2008A&A...484L...9W} that confirmed that the ejecta could be  collimated in polar directions producing a bipolar  morphology due to the latitude-dependent density stratification. A very similar phenomenon  is observed in the case of the symbiotic-like nova V407 Cygni \citep{2012MNRAS.419.2329O}. In the case of RS\,Oph and similar other cases, it is crucial to ascertain whether the model of a blast wave channeled by circumbinary material can lead to a structured bipolar nebula as observed by the HST \citep{2007ApJ...665L..63B} and also account for the evidence of fast jet-like outflow.

{\it Common envelope interactions} The Common Envelope (CE) phase occurs systematically when the binary is engulfed by the ejected material. It lasts longer and should naturally be more pronounced for slow novae which eject a larger amount of mass at velocities slower than that observed for fast novae.
The CE phase is expected to produce a density contrast between the equatorial (orbital plane) and polar directions \citep{1997MNRAS.284..137L}. The shaping occurs in a short time {\it after} the outburst of the nova, in comparison to the immediate effect of pre-existing circumbinary material.

{\it Magnetic fields} If the WD has a strong magnetic field (as in the case of polars), one can expect that this field may influence the dynamics and shaping of the material  ejected during the outburst. In the case of T\,Pyx, \citet{2011ATel.3782....1O} reported  the detection of photometric modulation at the orbital period during the tail of the eruption. The start time of the modulation (around day 149 after the start of the eruption) was close to the start of the X-ray plateau in the Swift X-ray light curve, when the region near the white dwarf has become unveiled. This was interpreted by the authors as strong evidence that the white dwarf has a very high magnetic field. Though these results are not formally available in a detailed publication as yet, one can conjecture whether the purported magnetic field in T Pyx finding is not partly responsible for the observations by \citet{2011A&A...534L..11C} which suggest an intrinsically bipolar ejection in T Pyx. In the parallel case of planetary nebulae, the role of magnetic fields in contributing to the formation of observed bipolar shapes  has often been invoked and discussed \citep{2002ARA&A..40..439B}; Sabin, Zijlstra and Greaves 2007 and references therein).

An additional important question that must be addressed  by observation is {\it to determine where the mass resides in a bipolar nebula}. Observational findings in this regard are often puzzling. Extreme examples are the bipolar nebula discovered around V445\,Pup \citep{2009ApJ...706..738W} apparently dominated by equatorial material, while the dust around V1280\,Sco seems to reside only in the caps (Chesneau et al., submitted).
The explosion of V445 Pup in 2000 occurred in helium-rich accreted material on the surface of a CO-type white dwarf \citep{2003A&A...409.1007A} with the pre-outburst light curve shape suggesting that the progenitor was a common-envelope binary \citep{2010PZ.....30....4G}. The strong CE environment that appears to have been present in the system is probably at the origin of the dense equatorial ejecta. It may be more difficult to explain bipolar shapes in which most of the mass resides in the caps. A similar difficulty applies also  to Planetary Nebulae where both situations are observed: 'caps-dominated' bipolar nebulae with tight waists such as Menzel\,3 or M2-9 \citep{2011A&A...527A.105L, 2007A&A...473L..29C, 2005AJ....129..969S} in contrast  with those exhibiting very dense and large equatorial torii such as NGC6302 \citep{2005MNRAS.359..383M}.

\section{Conclusions}

Optical interferometry has definitely a role to play in the study of novae. Currently limited by the sensitivity, only bright and slowly declining novae outbursts can be studied and pursued over a reasonable period of time,  but promising improvements are expected in the near-future. As an example, the observation of humps/superhumps and the study of dwarf novae, not mentioned in this review, should eventually be possible, at least when the source is at the maximum brightness state \citep{2000A&A...353..244H}. The increase of baselines and availability of medium resolution capabilities to resolve the broad emission lines should provide key insights into the shape, shaping  and kinematics of the ejecta at the earliest time.

%------------------------------------------------------------------------------%
%\section*{Acknowledgements}

%Here is where acknowledgements go, in an un-numbered section (which is
%specified by using the \verb|\section*{Acknowledgements}| command).

%------------------------------------------------------------------------------%
% bibliography: produced from ADS using custom format of                       %
%                                                                              %
%     %z132 \\bibitem[%\2%(y)%\3m]%{R}\n   %\8.1g,%\Y,%\q,%\V,%\ p             %
%------------------------------------------------------------------------------%

%------------------------------------------------------------------------------%
\end{document}